\documentclass[12pt,preprint]{aastex}

\newdimen\minuswidth    
\setbox0=\hbox{$-$}
\minuswidth=\wd0
\catcode`@=\active
\def@{\kern\minuswidth}
\newdimen\digitwidth    
\setbox0=\hbox{\rm0}
\digitwidth=\wd0
\catcode`!=\active
\def!{\kern\digitwidth}

\def\reference#1\par{\parindent0pt\hangindent20pt\hangafter1 #1\par}
\def\deg{^\circ}

\received{}
\begin{document}
\shorttitle{IR spectra of giant M stars}
\shortauthors{Rich \& Origlia}

\title{The First Detailed Abundances for M giants in the inner Bulge from Infrared Spectroscopy}
\altaffiltext{1}{
Data presented herein were obtained
at the W.M.Keck Observatory, which is operated as a scientific partnership
among the California Institute of Technology, the University of California,
and the National Aeronautics and Space Administration.
The Observatory was made possible by the generous financial support of the
W.M. Keck Foundation.}

\author{R. Michael Rich}
\affil{Physics and Astronomy Bldg,
430 Portola Plaza  Box 951547
Department of Physics and Astronomy, University of California
at Los Angeles, Los Angeles, CA 90095-1547\\
rmr@astro.ucla.edu}
\author{Livia Origlia}
\affil{INAF -- Osservatorio Astronomico di Bologna,
Via Ranzani 1, I--40127 Bologna, Italy,\\
livia.origlia@oabo.inaf.it}
\author{Elena Valenti}
\affil{ESO -- European Southern Observatory,      
Alonso de Cordoba 3107, Santaigo, Chile,\\
evalenti@eso.org}


\begin{abstract}
We report the first abundance analysis of 17 M giant stars in the inner Galactic
bulge at $(l,b)=(0\deg,-1\deg)$, based on $R=25,000$ infrared spectroscopy  $(1.5-1.8 ~\mu \rm m)$ using NIRSPEC at the Keck telescope.   Based on their luminosities and radial velocities, we identify these stars with a stellar population older than 1 Gyr.
We find the iron abundance 
$\rm \langle [Fe/H] \rangle = -0.22 \pm 0.03$,
with a $1\sigma$ dispersion of $0.14\pm 0.024$.
We also find the bulge stars have enhanced [$\rm alpha$/Fe] abundance ratio at the level
of +0.3 dex relative to the Solar stars, 
and low $\rm <^{12}C/^{13}C>\approx 6.5\pm 0.3$.   
The derived iron abundance and composition for this inner bulge
sample is indistinguishable from that of a sample of 
M giants our team has previously studied in Baade's Window $(l,b)=(0.9,-4)$.
We find no evidence of any major iron abundance or abundance ratio gradient between this inner field and Baade's window.
\end{abstract}


\section{Introduction}

\label{intro}

\begin{figure}
\plotone{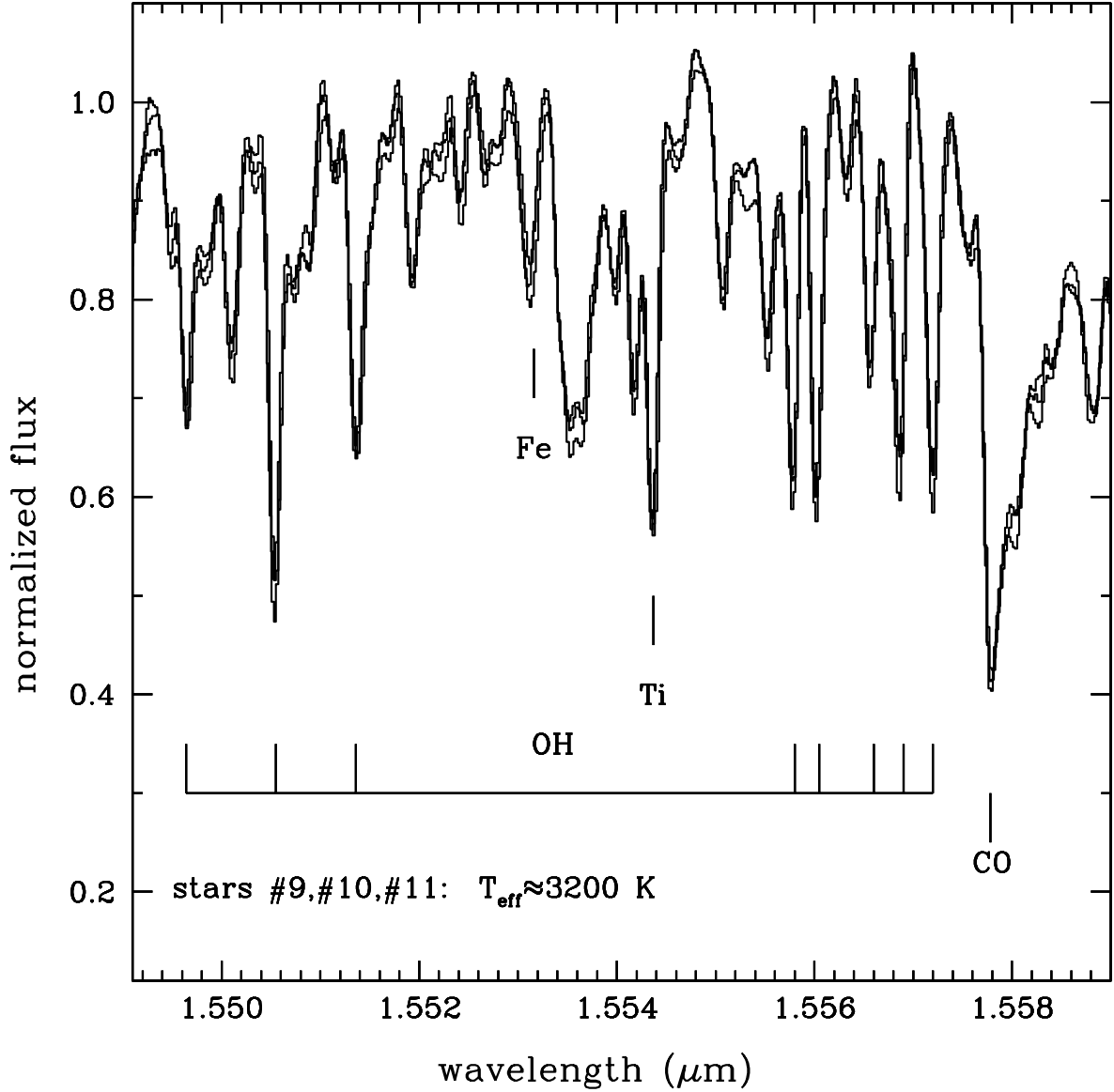}
\caption { NIRSPEC H--band spectrum showing the region near 1.555~$\mu$m of
three (\#9, \#10 and \#11) of the coolest and more metal rich stars in our sample.
A few major atomic lines and molecular bands of interest are
flagged. \label{spec}}
\end{figure}

High resolution infrared (IR) spectroscopy now enables the possibility of undertaking abundance analysis for cool 
late-type giants.  The technique is especially powerful in extending abundance studies to regions of high extinction, 
such as the inner disk and bulge.
We have reported the first detailed abundances for Galactic bulge M giants \citep{rich05} in Baade's Window, 
a field with low enough extinction that optical high resolution spectroscopy is also feasible 
\citep{MWR94,fmwr06,fmwr07,zoc06}.
Our IR spectroscopy has been in basically good agreement with these optical studies, finding roughly Solar mean 
abundance and enhanced alpha elements, 
consistent with a predominant chemical enrichment by type II SNe on a relatively short timescale 
\citep{mrm99,bal07}.

Here we exploit these techniques to push into the high extinction regions of the inner Galactic bulge, 
where optical spectroscopy is impossible.   
Our goal is to constrain whether an abundance or abundance ratio gradient is present in the bulge population.   
The Galactic Center contains old stars \citep[e.g.][]{fig04} and
the bulge in Baade's Window and other fields has been demonstrated to be globular cluster age
\citep{ort95,kui02,zoc03}.
Our aim has been to select the {\it old} red giant stars that lies near the red giant branch (RGB) tip 
a factor of four closer to the nucleus than the well studied Baade's window field, 
at $(l,b)=(0\deg,-1\deg)$, i.e. only
$\approx 140$ pc distant from the nucleus (adopting $R_0=8$ kpc),   
with the goal of measuring possible abundance gradients in the inner 
Galactic bulge.  

At present, there is no secure measurement of the bulge abundance gradient.   
\citet{fro99} used broadband JK photometry to probe fields along the major and minor axis
from $-4\deg<l,b<0\deg$, reaching within 0.2$\deg$ of the Galactic Center.  
This study, which found no indication of a significant gradient,
provided the input sample for \citet{ram00}, 
who also found no gradient using low resolution measurements of Ca, Na, and CO features in the 2$\rm \mu m$ window.    
The \citet{ram00} 
abundances were derived from a globular cluster-based calibration and
consequently their abundances for [Fe/H]$>$0 are an extrapolation; however,
that was unlikely to have altered their findings.  More recently,
\citet{zoc03} find a photometric abundance distribution at $(l,b)=(0.3\deg,-6.2\deg)$ very similar to that of 
\citet{fmwr06}  in Baade's window.
\citet{vie06} finds the same photometric abundance distribution in a proper motion selected sample 
of K giants in the Plaut field, at $b=-8\deg$.   
These studies suggest that the bulge does not have a measurable abundance gradient 
over its inner kpc, but they all rely on photometric abundance constraints 
(RGB color) rather than on spectroscopy.    

\section{Observations and abundance analysis}

A sample of M giants near the RGB tip
in the inner Bulge field at $(l,b)=(0\deg,-1\deg)$ 
has been selected from the 2MASS K, J-K color magnitude 
diagram, corrected for reddening by using the interstellar extinction maps of \citet{sch99}. 
Spectra for 17 stars with bolometric magnitudes $\rm M_{bol}\ge -4$  were taken 
during 3 observing runs on April 2005, May 2005 and May 2006, by using NIRSPEC \citep{ml98} at 
Keck II.
A slit width of $0\farcs43$  
giving an overall spectral resolution R=25,000,
and the standard NIRSPEC-5 setting, which
covers most of the 1.5-1.8 $\mu$m H-band
have been selected. 

The raw stellar spectra have been reduced using the
REDSPEC IDL-based package written
at the UCLA IR Laboratory.
Each order has been
sky subtracted by using nodding pairs and flat-field corrected.
Wavelength calibration has been performed using arc lamps and a second order
polynomial solution, while telluric features have been removed by using
a O-star featureless spectrum.
The signal to noise ratio of the final spectra is $\ge$30. 
Fig.~\ref{spec} shows
an example of the observed spectra for 3 among the coolest and most metal rich giants in our sample. 

A grid of suitable synthetic spectra
of giant stars has been computed 
by varying the photospheric parameters and the
element abundances, using an updated
version of the code described in \citet{OMO93}.
By combining full spectral synthesis analysis with equivalent widths
measurements of selected lines,
we derive  abundances for Fe C, O and 
other alpha-elements (Mg, Si, Ca and Ti).  The lines and analysis method are
described in \citet{orc02,ori04}
and subjected to rigorous tests in our previous studies
of  Galactic bulge field and cluster giants \citep[see][ and references therein]{ori04,rich05}.
Reference solar abundances are from \citet{gv98}.

In the first iteration, we estimate stellar temperature
from the $\rm (J-K)_0$ colors (see Table~\ref{tab1})
by using an average redddening of E(B-V)=2.9 \citep{sch99}
and the color-temperature transformation
of \citet{MFFO98} specifically calibrated on globular cluster giants.
Gravity has been estimated from theoretical evolutionary tracks,
according to the location of the stars on the Red Giant Branch (RGB)
\citep[see][and references therein for a more detailed
discussion]{ori97}.
An average value
$\xi$=2.0 km/s has been adopted for the microturbulence
\citep[see also][]{ori97}.
More stringent constraints on the stellar parameters are obtained by the
simultaneous spectral fitting of the several CO and OH molecular bands,
which are very sensitive to temperature, gravity and microturbulence variations 
(see Figs. 6,7 of \citet{orc02}). 

CO and OH in particular, are extremely sensitive to $T_{eff}$ in the range 
 3500 to 4500 K.
Indeed, temperature sets the fraction of molecular {\it versus} atomic
carbon and oxygen.
At temperatures $\ge$4500 K molecules barely survive; most of the
carbon and oxygen are in atomic form and the CO and OH spectral features
become very weak.
On the contrary, at temperatures $\le$3500 K most of the carbon and oxygen
are in molecular form, drastically reducing the dependence of the CO and OH
band strengths and equivalent widths on the temperature itself
\citep{ori97}. 

The final values of our best-fit abundances together with random errors
are listed in Table~\ref{tab1}.

As a further check on the statistical significance of our best-fit solution,
we also compute synthetic spectra with
$\rm \Delta T_{eff}=\pm$200~K, $\rm \Delta log~g=\pm$0.5~dex and
$\rm \Delta \xi=\mp$0.5~km~s$^{-1}$, and with corresponding simultaneous variations
of the C and O abundances (on average, $\pm$0.2~dex) to reproduce the depth of the 
molecular features.
As a figure of merit of the statistical test, we adopt
the difference between the model and the observed spectrum (hereafter $\delta$).
In order to quantify systematic discrepancies, this parameter is
more powerful than the classical $\chi ^2$ test, which is instead
equally sensitive to {\em random} and {\em systematic} scatters
\citep[see also][]{ori03,ori04}.
Our best fit solutions always show $>$90\% probability 
to be representative of the observed spectra, while 
those with $\pm$0.1~dex are significant at $\ge$1 $\sigma$ level.
Spectral fitting solutions with abundance variations of $\pm$0.2~dex,
due to possible systematic uncertainties of $\pm$200~K in temperature,
$\pm$0.5~dex in gravity or $\mp$0.5 km/s in
microturbulence have $<$30\% probability of being
statistically significant.
Hence, as a conservative estimate of the systematic error in the derived 
best-fit abundances,
due to the residual uncertainty in the adopted stellar parameters, one can
assume a value of $\le \pm 0.1$~dex.
However, it must be noted that since the stellar features under consideration 
show a similar trend with variations in the stellar parameters, although with different
sensitivities, {\it relative } abundances are less
dependent on the adopted stellar parameters (i.e. on the systematic errors)
and their values are well constrained down to $\approx \pm$0.1~dex
(see also Table~\ref{tab1}).

\section{Results and Discussion}

\begin{deluxetable}{lllclcrlllllllc}
\tabletypesize{\scriptsize}
\tablecaption{Stellar parameters and abundances for our sample of 
giant stars in the inner Bulge field at $(l,b)=(0,-1)$. \label{tab1}}
\tablewidth{0pt}
\tablehead{
\colhead{\#} & 
\colhead{RA}&         
\colhead{Dec}&         
\colhead{$\rm (J-K)^a_0$}&         
\colhead{$\rm T_{eff}$}&
\colhead{log~g}&
\colhead{$\rm v_r^b$}&
\colhead{$\rm [Fe/H]$}& 
\colhead{$\rm [O/Fe]$}& 
\colhead{$\rm [Si/Fe]$}& 
\colhead{$\rm [Mg/Fe]$}& 
\colhead{$\rm [Ca/Fe]$}& 
\colhead{$\rm [Ti/Fe]$}& 
\colhead{$\rm [C/Fe]$}& 
\colhead{$\rm ^{12}C/^{13}C$} 
}
\startdata
\multicolumn{15}{l}{}\\
1&17:49:53.6 &-29:31:22.4& 0.85 &4000& 1.0 & -327  & +0.02&  +0.28&  +0.28&  +0.30&  +0.28&  +0.38&  -0.32& 5.0\\
  &&& &&& & $\pm$ 0.11&  $\pm$ 0.16&  $\pm$ 0.20&  $\pm$ 0.16&  $\pm$ 0.19&  $\pm$ 0.20&  $\pm$ 0.17& $\pm$1.2\\
2&17:49:43.2 &-29:18:50.9& 1.25 &3400& 0.5 &  -66  & -0.40&  +0.41&  +0.30&  +0.32&  +0.30&  +0.23&  -0.40& 6.3\\
  &&& &&& & $\pm$ 0.08&  $\pm$ 0.11&  $\pm$ 0.21&  $\pm$ 0.09&  $\pm$ 0.12&  $\pm$ 0.14&  $\pm$ 0.11& $\pm$1.5\\
3&17:49:53.6 &-29:30:41.2& 1.04 &3600& 0.5 &  -76  & -0.23&  +0.22&  +0.33&  +0.32&  +0.33&  +0.33&  -0.22& 7.9\\
  &&& &&& & $\pm$ 0.08&  $\pm$ 0.11&  $\pm$ 0.17&  $\pm$ 0.08&  $\pm$ 0.12&  $\pm$ 0.14&  $\pm$ 0.10& $\pm$2.0\\
4&17:49:46.5 &-29:19:17.8& 1.37 &3200& 0.5 &  -7   & -0.11&  +0.32&  +0.21&  +0.30&  +0.31&  +0.22&  -0.29& 6.3\\
  &&& &&& & $\pm$ 0.07&  $\pm$ 0.11&  $\pm$ 0.16&  $\pm$ 0.08&  $\pm$ 0.12&  $\pm$ 0.13&  $\pm$ 0.10& $\pm$1.5\\
5&17:49:51.2 &-29:31:00.4& 1.21 &3400& 0.5 & -192  & -0.20&  +0.22&  +0.25&  +0.24&  +0.30&  +0.18&  -0.25& 6.3\\
  &&& &&& & $\pm$ 0.08&  $\pm$ 0.12&  $\pm$ 0.17&  $\pm$ 0.09&  $\pm$ 0.12&  $\pm$ 0.15&  $\pm$ 0.11& $\pm$1.5\\
6&17:49:36.7&-29:18:53.8& 1.40 &3200& 0.5 &  -32  & -0.40&  +0.27&  +0.24&  +0.27&   +0.30&  +0.25&  -0.30& 5.0\\
  &&& &&& & $\pm$ 0.08&  $\pm$ 0.11&  $\pm$ 0.14&  $\pm$ 0.09&  $\pm$ 0.12&  $\pm$ 0.14&  $\pm$ 0.11& $\pm$1.2\\
7&17:49:40.8 &-29:18:17.2& 1.09 &3600& 0.5 &  +45  & -0.24&  +0.30&  +0.34&  +0.34&  +0.34&  +0.34&  -0.26& 5.0\\
  &&& &&& & $\pm$ 0.03&  $\pm$ 0.05&  $\pm$ 0.13&  $\pm$ 0.04&  $\pm$ 0.05&  $\pm$ 0.07&  $\pm$ 0.08& $\pm$1.2\\
8&17:49:53.3 &-29:30:28.02& 1.10 &3600& 0.5 &  -1   & -0.17&  +0.26&  +0.27&  +0.29&  +0.27&  +0.35&  -0.33& 6.3\\
  &&& &&& & $\pm$ 0.03&  $\pm$ 0.05&  $\pm$ 0.13&  $\pm$ 0.04&  $\pm$ 0.05&  $\pm$ 0.07&  $\pm$ 0.08& $\pm$1.5\\
9&17:49:47.0 &-29:18:49.2& 1.47 &3200& 0.5 &  -15  & -0.10&  +0.27&  +0.22&  +0.29&  +0.30&  +0.23&  -0.30& 7.9\\
  &&& &&& & $\pm$ 0.04&  $\pm$ 0.06&  $\pm$ 0.13&  $\pm$ 0.04&  $\pm$ 0.06&  $\pm$ 0.07&  $\pm$ 0.08& $\pm$2.0\\
10&17:49:47.6 &-29:18:59.8& 1.42 &3200& 0.5 & +121  & -0.17&  +0.19&  +0.22&  +0.26&  +0.27&  +0.27&  -0.23& 7.9\\
  &&& &&& & $\pm$ 0.04&  $\pm$ 0.06&  $\pm$ 0.13&  $\pm$ 0.04&  $\pm$ 0.06&  $\pm$ 0.07&  $\pm$ 0.08& $\pm$2.0\\
11&17:49:35.5 &-29:18:46.1& 1.49 &3200& 0.5 & -168  & -0.07&  +0.15&  +0.17&  +0.24&  +0.27&  +0.22&  -0.33& 6.3\\
  &&& &&& & $\pm$ 0.04&  $\pm$ 0.06&  $\pm$ 0.13&  $\pm$ 0.04&  $\pm$ 0.06&  $\pm$ 0.07&  $\pm$ 0.08& $\pm$1.5\\
12&17:49:41.8 &-29:18:25.4& 1.36 &3200& 0.5 &  -94  & -0.10&  +0.16&  +0.20&  +0.29&  +0.30&  +0.30&  -0.25& 7.9\\
  &&& &&& & $\pm$ 0.09&  $\pm$ 0.14&  $\pm$ 0.17&  $\pm$ 0.10&  $\pm$ 0.13&  $\pm$ 0.15&  $\pm$ 0.11& $\pm$2.0\\
13&17:49:39.5 &-29:20:12.3& 1.00 &3800& 0.5 & +165  & -0.34&  +0.43&  +0.44&  +0.41&  +0.34&  +0.40&  -0.26& 6.3\\
  &&& &&& & $\pm$ 0.09&  $\pm$ 0.10&  $\pm$ 0.21&  $\pm$ 0.10&  $\pm$ 0.13&  $\pm$ 0.14&  $\pm$ 0.12& $\pm$1.5\\
14&17:49:49.0 &-29:19:50.8& 1.07 &3400& 0.5 & -181  & -0.11&  +0.27&  +0.26&  +0.27&  +0.31&  +0.19&  -0.29& 7.9\\
  &&& &&& & $\pm$ 0.07&  $\pm$ 0.11&  $\pm$ 0.19&  $\pm$ 0.08&  $\pm$ 0.12&  $\pm$ 0.15&  $\pm$ 0.10& $\pm$2.0\\
15&17:49:39.4 &-29:20:21.3& 1.33 &3400& 0.5 &  -34  & -0.42&   0.37&   0.32&   0.32&   0.32&   0.26&  -0.38& 6.3\\
  &&& &&& & $\pm$ 0.08&  $\pm$ 0.11&  $\pm$ 0.15&  $\pm$ 0.09&  $\pm$ 0.12&  $\pm$ 0.14&  $\pm$ 0.11& $\pm$1.5\\
16&17:49:40.7 &-29:20:13.3& 1.35 &3600& 0.5 &  -90  & -0.46&   0.37&   0.36&   0.34&   0.36&   0.36&  -0.24& 6.3\\
  &&& &&& & $\pm$ 0.08&  $\pm$ 0.10&  $\pm$ 0.15&  $\pm$ 0.09&  $\pm$ 0.12&  $\pm$ 0.14&  $\pm$ 0.11& $\pm$1.5\\
17&17:49:35.4 &-29:17:45.0& 1.25 &3400& 0.5 & -206  & -0.25&  +0.22&  +0.19&  +0.23&  +0.25&  +0.25&  -0.35& 5.0\\
  &&& &&& & $\pm$ 0.08&  $\pm$ 0.11&  $\pm$ 0.17&  $\pm$ 0.09&  $\pm$ 0.12&  $\pm$ 0.14&  $\pm$ 0.10& $\pm$1.2\\
\enddata                       
\tablenotetext{a}{The (J--K) colors are from 2MASS and have been corrected 
for reddening using E(B-V)=2.9.} 
\tablenotetext{b}{Heliocentric radial velocity in $\rm km~s^{-1}$.}
\end{deluxetable}

\begin{figure}
\plotone{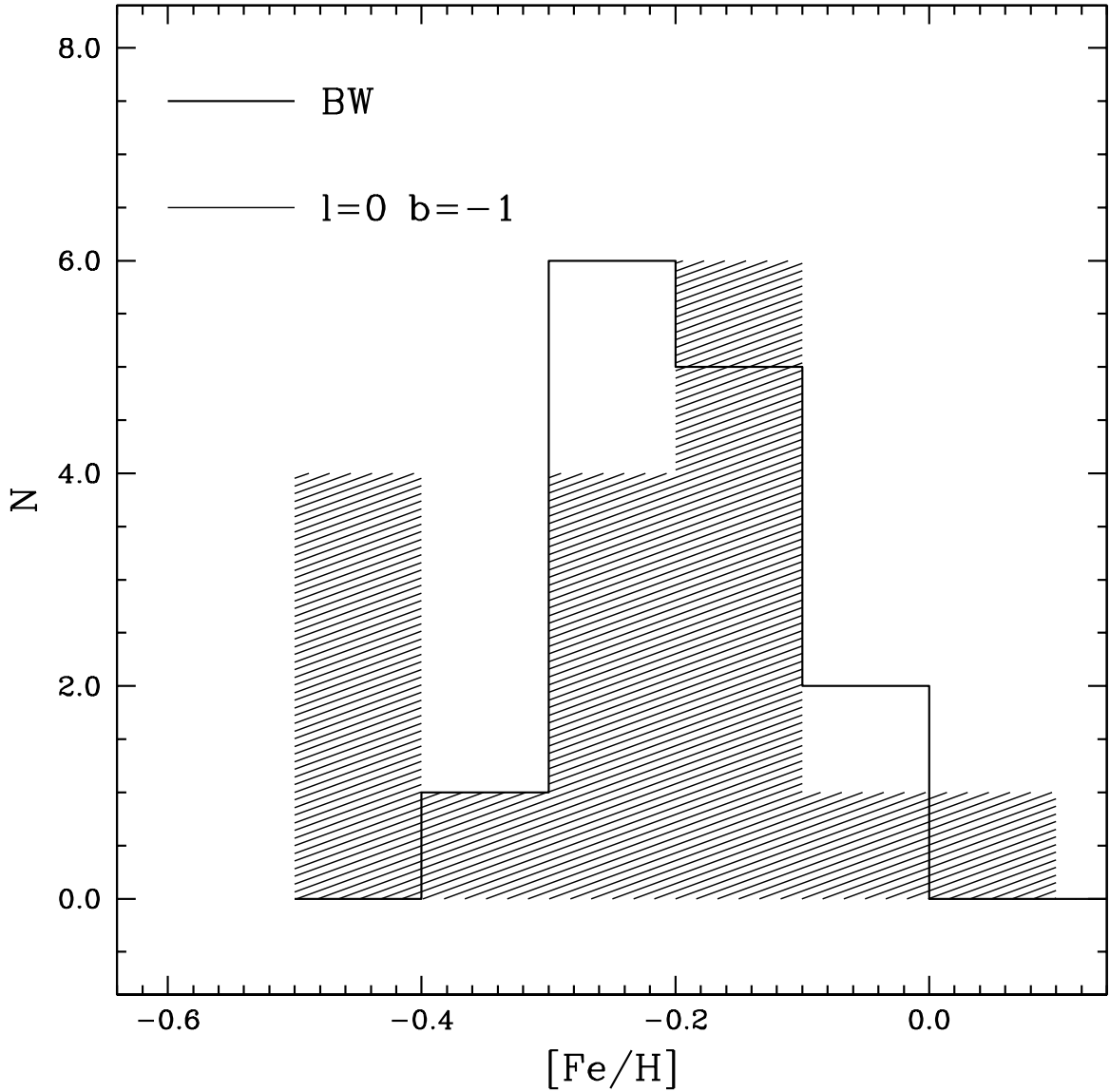}
\caption {Histogram of the metallicity distribution in [Fe/H],
for the observed giants in the inner bulge field at $(l,b)=(0\deg,-1\deg)$ (shaded histogram) 
and in Baade's window $(l,b)=(0.9\deg ,-4\deg)$ (solid line, \citet{rich05}), for comparison.  Notice that
there is no significant difference between the two fields.
\label{histo}}
\end{figure}

\begin{figure}
\plotone{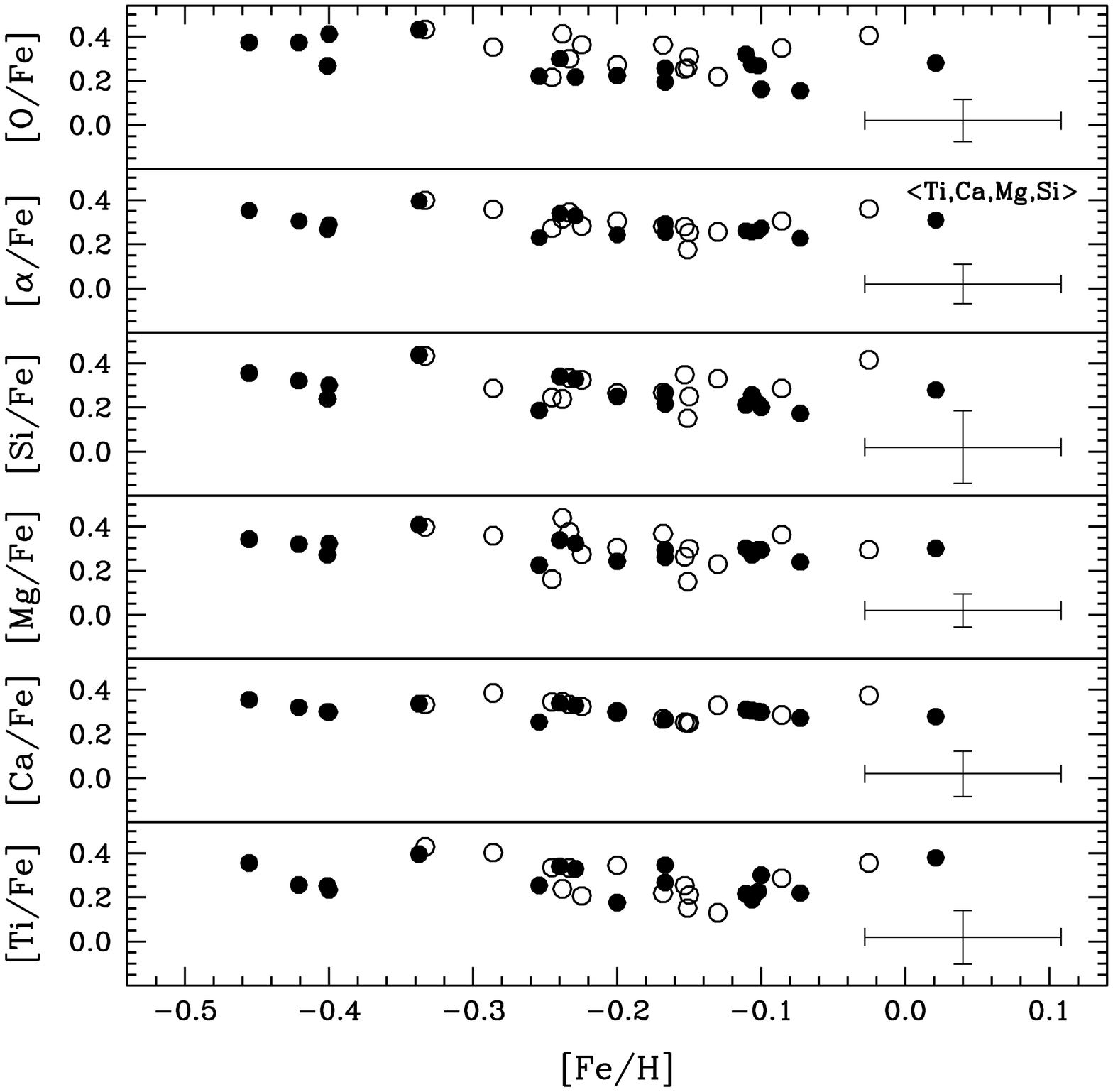}
\caption{$\alpha/Fe$ 
as a function of [Fe/H] for the observed giants in the inner bulge field at $(l,b)=(0\deg ,-1\deg )$ (filled circles) 
and in Baade's window (open circles, \citet{rich05}), for comparison.
Typical errors are plotted in the bottom-right corner of each panel.  
There is no evidence for a gradient in $[\alpha/Fe]$ in the bulge.
\label{alpha}}
\end{figure}

For the observed M giants we measure average $\rm <[Fe/H]>=-0.22 \pm 0.03$ 
and dispersion $1\sigma=0.140\pm 0.024$.
This [Fe/H] abundance distribution is very similar 
to the one measured in the Baade's window \citep{rich05}, with almost identical 
average value within the error and only marginally greater dispersion. 

We also confirm 
the lack of both {\it metal poor} and {\it super metal rich} M giants.  This
was originally noted in the \citet{rich05} Baade's Window M giant
distribution, but now persists in a sample that, in total, is nearly three
times as large.
We know that the metal poor end will be poorly represented, as few stars with [Fe/H]$<-1$ will evolve 
to be cool enough to develop TiO bands and become M giants at the RGB tip.  
However, 
the lack of {\it super metal rich}
M giants is a concern. 
Indeed, although they represent only a fraction ($\le $20\%, \citet{fmwr06}) of the bulge population, 
they are definitely present in all bulge K giant studies.  We speculate that their low intrinsic numbers and rapid evolution
may make their detection statistically unlikely.
Moreover, the possible enhanced mass loss 
at super solar metallicities \citep[see e.g.][]{cas93} 
may also prevent the most metal rich stars from reaching the RGB tip.  
It is important to explore these issues with a larger sample, but 
the total of 31 M giants observed in these two fields is now large enough
that the lack of stars with [Fe/H]$>$0.1 is a significant concern.

Our survey shows a rather homogeneous $\alpha$-enhancement by $\approx$+0.3$\pm 0.1$ dex
up to solar metallicity, without significant differences among the various
$\alpha$-elements (see Fig.~\ref{alpha}), and similar to measurements in the Baade's window.  
Such an enhancement of the [$\alpha$/Fe] abundance ratio suggests  
that at the epoch of the bulge formation, mainly type II SNe were contributing to the enrichment 
of the interstellar medium. It also indicates that the overall formation process should have ended before 
the bulk of type Ia SNe have exploded, i.e. well within a few Gyr, the precise timescale depending 
on the SN modeling \citep[see e.g.][]{gre07}.

We have also derived carbon abundances for all of the observed stars from
analysis of the CO bandheads.
We find some degree of [C/Fe] depletion ($\rm <[C/Fe]>=-0.29 \pm 0.02$) with respect to solar values
and very low $^{12}$C/$^{13}$C ($\rm <^{12}C/^{13}C>=6.5 \pm 0.3$) , as
also measured in Baade's window M giants, 
confirming the
presence of extra-mixing processes during the RGB phase of evolution in metal-rich environments.

From our detailed analysis of chemical abundances and abundance patterns we 
can thus exclude any major abundance gradient in the inner bulge.

Finally, we have also measured radial velocities and velocity dispersion for the observed stars,
as for the Baade's window giants.
The overall velocity dispersion of the global sample of 31 M giants 
turns out to be $\rm \sigma_{v_r}\approx 116 \pm 15$ km/s,
fully consistent with bulge kinematics.

Theory concerning the formation of spheroids has evolved, beginning with 
\citet{els}, who proposed violent relaxation and dissipative collapse.   
In the CDM scenario, spheroids would form from the mergers of disks.
Further, large galaxy surveys now suggest that the spheroid-dominated 
"red sequence" has grown in mass since $z\approx 1$ \citep{bell04}, 
implying that some spheroids have come into existence relatively later than the age of 
the Galactic bulge population.   
A widely supported scenario for bulge formation was initially proposed by \citet{raha91}.   
In this scenario, a massive disk develops bending modes that 
ultimately thicken the disk vertically, producing a peanut-shaped bar.  
These ideas have been developed by many other theoretical studies e.g. \citet{pfenn90}.  
The bar thickening hypothesis predicts that no vertical abundance gradient should be present.
Carefully weighing the existing evidence, \citet{kor04}
conclude that the Galactic bulge has likely been formed via secular processes, 
but they are perplexed by the strong evidence for its age being similar to that of the halo.   
The recent survey of bulge kinematics by \citet{brava07} supports the idea that the bulge is 
kinematically distinct from the inner disk and that while the deprojection of the bulge demands 
some elongated structure \citep{zhao96}, the new velocity field also requires the presence of 
retrograde orbits by
construction.   While the lack of an abundance gradient is consistent with secular
evolution models, the great age of the bulge/bar remains as a problem for this picture.

\acknowledgments
R. Michael Rich acknowledges support from 
grants AST-039731 and
AST-0709479 from the National Science Foundation.
Livia Origlia acknowledges financial support by the 
Ministero dell'Istru\-zio\-ne, Universit\`a e Ricerca (MIUR) and the 
Istitututo Nazionale di Astrofisica.  We are grateful to Ian Mclean and
the UCLA Infrared laboratory for the construction of NIRSPEC, and
to the staff of the W. M. Keck observatory for support during the observations.

\end{document}